\documentclass[]{aa}

\def\be{\begin{equation}}
\def\ee{\end{equation}}
\def\bea{\begin{eqnarray}}
\def\eea{\end{eqnarray}}

\renewcommand{\thesection}{\arabic{section}}
\renewcommand{\thesubsection}{\arabic{section} \arabic{subsection}}

\bibpunct{(}{)}{;}{a}{}{,}

\usepackage[varg]{txfonts}
\usepackage{graphicx}
\usepackage{caption}
\usepackage{subcaption}
\usepackage{amssymb}
\usepackage{enumitem}
\usepackage{natbib}
\usepackage{amsmath}
\usepackage{hyperref}

\usepackage{color}

\begin{document}

\title{Acceleration of cosmic rays by double shock waves in galaxy clusters: application to radio relics}

\author{Grazyna Siemieniec--Ozieblo
\and Mariia Bilinska
}

 \institute {Astronomical Observatory, Jagiellonian University,  ul. Orla 171, 30--244 Krak\'ow, Poland\\
\email{grazyna@oa.uj.edu.pl; mariiaklochko@gmail.com}
}

\date{Received <date> /
        Accepted <date>}

\abstract{Radio relics in galaxy clusters are known to be good laboratories for verification of the applicability of the diffusive shock acceleration (DSA) model in its canonical version. The need for such verification stems from the inconsistencies in the shock properties resulting from radio observations compared to X-ray observations.
}
{
 In this article we aim to explore how the presence of a second shock in the vicinity of a relic modifies the spectrum of accelerated electrons and decipher which of the involved parameters can have a significant impact on their shape.
}
{We analytically studied DSA of cosmic rays in two stationary shocks aiming to investigate the change of the distribution function. The latter eventually leads to spectrum slope deviations visible in different observations and simulations that do not appear to be explained by the case wherein cosmic rays interact with a single shock wave.
}
{
We obtain a complex distribution function $f(x,p)$ depending on many parameters (distance between two shocks, compression ratios, spatial diffusion coefficients, injection value, etc.). This function reveals modifications that occur because of the coupled acceleration in both shocks. Apparently, deviations in the particle spectrum from the pure power law depend on a few parameters such as $Q_{1}/Q_{2} $, $\kappa_{1}/\kappa_{2} $, $r_{1}/r_{2} $, and L. Although we do not verify this idea by taking a particular cluster  as an example, we demonstrate a potential cause of spectral disturbances in radio relics. In general terms, our findings appear to correlate with results from the literature when the distance between the shocks is of the order of the width of a radio relic and $\kappa_{1}/\kappa_{2} \propto 3$.
}{}

\keywords{Galaxies: clusters: general --- acceleration of particles --- cosmic rays --- shock waves --- methods: analytical}

\maketitle

\section{Introduction}

According to the large-scale structure (LSS) formation paradigm, the evolution of galaxy clusters is a hierarchical process consisting of continuous mergers and accretion of smaller systems. Both acts of merging and accretion are accompanied by the appearance of shock waves. In the case of galaxy clusters, these are shocks moving at nonrelativistic velocities. When shocks are moving, their huge kinetic energy is distributed into different channels (e.g., \citealt{2009MNRAS.395.1333V}). Although the main role of shocks is to heat plasma in intra-cluster medium (ICM), clusters are also  energy reservoirs for a number of other nonthermal phenomena leading on the one hand to particle acceleration and on the other to amplification of magnetic fields and magnetohydrodynamic (MHD) turbulence generation.

Our main astrophysical motivation is to explain the whole variety of nonthermal and thermal processes  occurring at different spatial scales of clusters in both central regions, where the interaction of active galactic nuclei (AGNs) with the cluster medium is of primary importance,  and other places, particularly at the periphery, with conditions favoring multi-frequency emissions.

Nonthermal radiation observed from galaxy clusters, which results from the interaction of nonthermal relativistic particles, that is, cosmic rays (CRs), with the other components of the cosmic plasma of the ICM,  has built-in characteristic features  that are specific to a given acceleration mechanism. The diffusive synchrotron radiation that we observe in clusters is a manifestation of the diffusive Fermi-type acceleration regime, as in the vast majority of nonthermal radiation emitted by different astrophysical sources. Thus, this is the basic motif of astrophysical nature, which 
also indicates galaxy clusters as relevant objects with which to test the Fermi acceleration mechanism of the first order. Features of the nonthermal diffusion emission produced in the ICM, in particular its deviation from a simple, so-called universal power-law spectrum, provides the basis for  verification of the DSA model in this environment. In light of recent observations, radio relics are particularly good candidates for such verification procedures (e.g., \citealt{2019SSRv..215...16V}, \citealt{2020MNRAS.493.2306B}, \citealt{2019SSRv..215...14B}). However, a multi-messenger approach used to study relics in both the radio and X-ray domains shows significant discrepancies, especially in estimations of shock-wave characteristics \citep{2013MNRAS.433..812O,2016ApJ...818..204V}.

Diffusive radio emission of cluster relics occurring typically on the edges of the clusters of galaxies is associated, because of the lack of an obvious optical counterpart, with the ICM environment and indicates the presence of relativistic electrons with $ \mathrm{\mathit{\gamma} \approx 10^{4}}$ which are being accelerated within the hot $ \mathrm{T \approx 5 - 10 \:keV} $ (where T is temperature) and magnetized  $\approx \mu$G plasma. On the other hand, there is a statistical correlation between the global features of relics, including X-ray and radio brightness, and the dynamic state of the cluster (see, e.g. \citealt{2002MNRAS.331.1011E}, \citealt{2016ApJ...831..110G}, \citealt{2017ApJ...838..110G}, \citealt{2019ApJS..240...39G} and \citealt{2019ApJ...882...69G}). And vice versa: only in dynamically disturbed clusters (current shock waves) can one  observe the diffuse radio emission in the form of a relic \citep{2019SSRv..215...16V}.
Therefore, particularly for radio relics, where there is no controversy as to the relationship between the existence of a shock wave and a radio signal, reasonable verification or clarification of the DSA acceleration model can be afforded.

In practice, our aim is to obtain a consistent interpretation of the shock-wave features, which is usually identified with a merger shock, seen in observations in both the radio and X-ray domains. A good, dimensionless parameter characterizing the power of  a particular shock is the sonic Mach number $\mathrm{M = v_{sh}/v_{sound}}$. In the linear version of the DSA model, this number entirely determines the form of the spectrum of accelerated particles. Therefore, in a first approximation, the agreement obtained between two independent Mach numbers, that is, the number resulting from X-ray observations and that resulting from the radio spectrum, should be sufficient confirmation of the proper description of the diffusive radio emission of the relic by the DSA model \citep{2010ApJ...715.1143F,2012PASJ...64...67A,2016arXiv160607433S}. However, the comparison of relativistic electron features seen through the radio observations with the X-ray shock diagnostics implied by thermal plasma fraction reveals many discrepancies in the estimated shock characteristics.

The list of the most common incompatibilities is summarized by the following problems:
\begin{itemize}
        \item The Mach numbers resulting from the X-ray emission often disagree with
(are smaller than) those resulting from the radio spectral index \citep{2014MNRAS.445.1213S,2014MNRAS.441L..41S,2016ApJ...818..204V,2017MNRAS.471.1107H,2017A&A...600A.100A,2018ApJ...857...26H}.
        \item The relics radio spectrum often reveals deviations from the power-law spectrum  \citep{2015A&A...575A..45T,2016MNRAS.455.2402S,doi:10.1007/s10509-016-2844-7,2017A&A...600A..18K,2019arXiv191108904R}.
        \item Sometimes one can see a spatial offset in the locations of X-ray and radio shocks \citep{2013MNRAS.433..812O}.
        \item There is a lack of consistency between the deduced Mach number and the observed suitable electron acceleration efficiency \citep{2013ApJ...764...95K,2020A&A...634A..64B,2020arXiv200309825L}.
\end{itemize}   

In the literature, there has been a wide spectrum of ideas put forward to explain these inconsistencies. Most of the concepts explaining the problem of the above discrepancies require the presence of a residual electron population re-accelerated via DSA or refer to consideration of significant shock turbulence or the involvement of adiabatic compression of plasma lobes from old radio sources. Unfortunately, even these varied approaches do not comprehensively solve all the problems listed above.

In this article, we refer to the idea put forward by \citet{2016A&A...596A..30S}, according to which the phenomenon of radio relics could be the result of the simultaneous presence of two shock waves with different Mach numbers; one being a forward merger shock approaching the other, the two with definitively different characteristics, for example,  internal accretion shock or inflow shock. We expect that within such a scenario, the DSA acceleration process will be modified by the presence of this latter shock. In particular, the resulting morphology of radio emission, including the flattening or steepening of relic spectrum observed in certain
cases, may be the property linked to the mutual influence of two shocks situated near to one another. Such a two-shock
configuration has seldom been discussed in the literature. However, several papers have recently appeared  that consider a two-shock scenario in the context of either magnetosphere or solar environment (e.g., \citealt{doi:10.1029/2011JA016669}). Moreover, \citet{2014MNRAS.441L..41S} show that a single power law does not accurately describe the spectrum of different parts of radio relics in the Toothbrush and Sausage clusters. \citet{2016MNRAS.455.2402S} suggest a broken power-law model to describe the spectral steepening around 2-2.5 GHz of each of the relics. We believe that the observations of 1RXS J0603.3+4214 and CIZA J2242.8+5301 radio relics require theoretical evaluation also at much lower frequencies. At 120-150 MHz there is a clear concave, “saddle-like” feature which we describe here via our double-shock model.

Here, in galaxy clusters we consider the concept of such a "double" shock, that is, a strong outer accretion shock and a weak inner merger shock with cosmic rays (CRs) trapped in-between them. One might expect a higher level of turbulence in such inter-shock collapsing traps which could have different diffusivity and thus might explain the subtle features at the electron emission spectrum at $200-300$ MHz visible in both observations and simulations.

We focus here on the parameters that may be of significant importance with respect to changes in the profile of the electron spectrum. We describe the parameters, vary them, and try to explain the trend of the concave shape that appears in the  spectrum of cosmic ray electrons (CRe) and is revealed in both observations and simulations of shock waves in galaxy clusters (e.g., \citealt{2016A&A...591A.142B} and \citealt{doi:10.1186/s40623-018-0799-3}). To do this we derive a new solution for the adequate electron transport equation and analyze the influence of some of the most important dimensionless parameters on the shape of the  spectrum. We also explicitly show that the fossil electrons swept away by the inner shock could be one of the causes of the occurrence of spectral shape distortions. We show in general that the CRe spectrum is very susceptible to dependence on certain parameters, contrary to what is expected in pure DSA.

Thus, the purpose of this paper is to show that a double-shock build up with one merger shock and an invisible accretion (virial) shock can change the so-called universal CR spectrum even in the linear regime of acceleration. The outline of the article is as follows: in Sect. 2 we present the mathematical model that we rely on to obtain the results which we describe and discuss in Sect. 3. The results section is followed by a summary.

\section{Theoretical approach}

Cosmic-ray acceleration in a system consisting of two relatively close shocks was recently analyzed in the literature, mainly in the context of interplanetary shocks (e.g., \citealt{doi:10.1029/2011JA016669}). Both the empirical spectra of detected ions and the results of particle simulations showed that in the case of two shocks, the description of the acceleration process via DSA in terms only of a single shock does not reproduce the observed results. The spectral index of radio emission for a pair of close shocks turned out to be harder than that of a single shock.

In a recent paper, \citet{doi:10.1186/s40623-018-0799-3} discussed the results of a test-particle simulation describing particle acceleration in the presence of two shocks, which finally generated a bent spectrum of particles in the form of a double power law. Here, in the case of a radio relic we also examine the DSA process in a system including two shocks. In the current section we describe the theoretical model that we use in our calculations.

\begin{figure}[htb]
        \begin{center}
                \includegraphics[totalheight=5.5 cm]{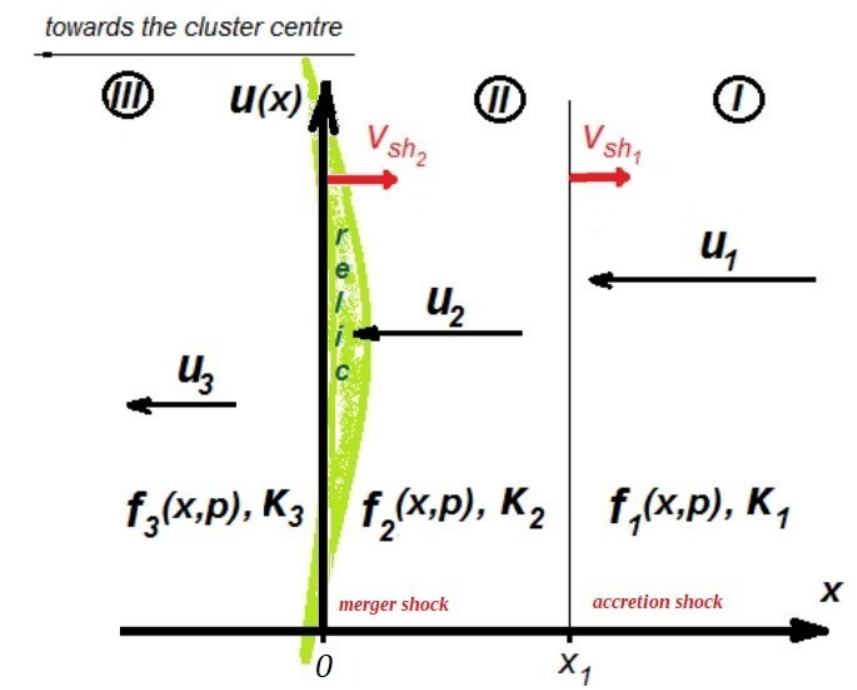}
        \end{center}
        \vspace{-4mm}
        \caption{Schematic image of a one-dimensional double-shock system. The image shows the  directions of two shock waves, the flow speeds, the position of a radio relic, and the three shock regions.}
        \label{Figure1}
\end{figure}

A merger shock wave propagating outward from the cluster is considered. The inflowing gas also offers an additional mechanism for shock formation within a galaxy cluster volume, which takes place both at the cluster border and within the whole volume of a galaxy cluster. Such an inflowing shock (hereafter referred to as an "accretion" shock) has a lifetime scale comparable to the Hubble time, while the lives of the merger shocks are of the order of the cluster dynamical timescale. Accretion shock allows matter to be accreted along the filaments into the ICM. In the current paper we consider a system of two such shocks: an accretion shock $sh1$ and a merger shock $sh2$. As the accretion shock is formed due to the inflow of matter from outside the cluster, it is stronger than the merger shock, that is, its Mach number is larger than the merger shock Mach number  $(\geq 4; \sim 2)$ \citep{2015ApJ...812...49H}. Compared to the merger shock, the accretion shock can be considered as stationary; moreover, in our consideration we treat both shocks as not moving with respect to each other, which means that the timescale of acceleration is less than the time needed for the merger shock to overlap the accretion shock. The position of the radio relic should roughly coincide with the merger shock. In our analysis we use the diffusion convection equation for three regions defined with respect to the positions of the  two shocks (see Fig.~\ref{Figure1}): an absolute downstream region, an inter-shock region (between the two shocks), and an absolute upstream region.

In each of the shock-related regions, values of diffusion coefficients $\kappa_{i}$, flow velocity $u_i,$ and particle distribution function $f_{i}(x,p)$ ($i=1,2,3\:$where indexes $1, 2, 3$ correspond to the absolute upstream, inter-shock, and absolute downstream regions, respectively) are different. We consider the injection to be stationary which makes the injection terms $Q_1$ and $Q_2$ constant.

As they spread towards the outskirts of the galaxy cluster, the presence of the shock waves leads to a decrease in the flow velocity in the directions of the cluster center; hence $u_{1}>u_{2}>u_{3}\:$and $x_1,\:$ $x_2\:$denote the positions of the shocks; furthermore (as seen in Fig.\ref{Figure1}), for simplicity we set $x_{2}=0$. The above parameters are the main transport parameters of the system. Additionally, we consider compressions $r_{1}=u_{1}/u_{2}$ and $r_{2}=u_{2}/u_{3}$ which are ratios of flow velocities.

In our consideration, both merger and accretion shocks are infinite and planar. However, these meet different mechanisms of injection: for the outer accretion shock it is a supra-thermal seed accompanied by a marginally small number of fossil electrons that could be able to diffusively reach that shock, while for the inner merger shock the presence of mildly relativistic electrons is crucial. The fossil population is assumed, for simplicity, to consist of mono-energetic (not power-law) electrons injected by AGNs to the ICM, cooled, and then swept away by merger shocks. The injection mechanisms still require a better understanding and detailed description. Nevertheless, recent results from simulations of the collision-less shocks show that the fraction of injected particles is a function of many parameters (e.g., ~\citealt{2018JPlPh..84c7101C}). Therefore,  it is impossible to quantify the $Q_{i}$ values  in general
terms. In the following, we use the nondimensional parameter $Q_{2}/Q_{1}$ to represent the ratio of both injections. To estimate the range of the ratio of the injection amplitudes $Q_2/Q_1$, we roughly assume the proportionality relation to the particle densities injected into the acceleration process, i.e., $Q_2/Q_1 \propto n_{e,2}/n_{e_{th},1}$ where $n_{e,2}$ changes from $n_{e_{th},2}$ in case of supra-thermal injection at the merger shock up to $n_{e_{rel},2}$ for injection occurring in the lobe environment which could be placed there. Here, $n_{e}$ is the number density of injected electrons; $n_{e_{th}}$ are thermal electrons and $n_{e_{rel}}$ are relativistic. Due to the larger gas density at shock 2 and more efficient injection in the case of the compressed (e.g., radio galaxy) lobe, we expect that the difference between the number of particles injected into both shocks may become essential (see Sect. 3.1) for the acceleration efficiency of this double-shock system.

Therefore, considering a rather simple model and focusing mainly on the role of injection in this dual-shock system, we consider the diffusion coefficients to be independent of the energy of the particles, but different in each of the regions: $\kappa_{i}\neq\kappa_{i}(p)$. Otherwise, the results of a diffusion-based mechanism of acceleration become dependent on the "momentum scale". If the diffusion coefficient were to show a dependence on momentum (i.e., $\lambda (p)$), one might expect the kind of "spectral breaking" feature. This could be produced because of the momentum dependence in later discussed, dimensionless parameter: shock distance/diffusion length(p). This shows that only particles with higher momenta than the "breaking" point value would be accelerated in both shocks. To avoid the description of such a complex scenario we restrict ourselves to the above-described, unrealistically simple form of diffusion coefficient.

In such a system of two relatively close shocks, particle distribution is governed by a diffusion-convection equation. The CR transport equation for the space distribution function is as follows:

\begin{multline}
    u_{i}(x) \frac{\partial f_{i}(x,p)}{\partial x}-\kappa_{i}(x) \frac{\partial^{2} f_{i}(x,p)}{\partial x^2}
    -\frac{1}{3} \frac{\partial u_{i}(x)}{\partial x}p\frac{\partial f_{i}(x,p)}{\partial p}-Q_{i}(x,p)=0,
    \label{eq:01}
\end{multline}
where $f(x,p)$ is the coordinate- and momentum-dependent distribution function; $u_{i}$  is the velocity of inflowing matter ($u_{1}>u_{2}>u_{3}$); $\kappa_{i}(x)$ is the diffusion coefficient; and $\mathrm{Q_{i}(x,p)=Q_{i} \:\: \delta(x-x_{i})\delta(p-p_{0i})}$ is the source term at the position of each of the two shocks.

Below, we assume flow velocities and diffusion coefficients in all three regions to be constant for each region, i.e., $u_{i}\ne u_{i}(x)$ and $\kappa_{i} \ne \kappa_{i}(x).$ As discussed above, we consider a stationary scenario with $f_{i} \ne f_{i}(t).$

Under these assumptions we are searching for the separable solution, $f(x,p)=g(x) \: h(p),$ of eq.~(\ref{eq:01}), leading to the general solution of the form (for $i=1,2$)

\begin{equation}
f_{i}(x,p)=\frac{\kappa_{i}}{u_{i}}A_{i}(p)\exp{\Biggl(-\frac{u_{i}}{\kappa_{i}}x\Biggr)} + B_{i}(p)
\label{eq:02}
.\end{equation}

Boundary conditions determine whether this separable solution is relevant or not. Below we postulate the physically reasonable boundary conditions:

\begin{itemize}[label=$\bullet$]
        \item $f_{1}(\infty, p)=0;$
        \item $f_{1}(x_{1}, p)=f_{2}(x_{1}, p)$ and $f_{2}(x_{2}, p)=f_{3}(x_{2}, p);$
        \item $\mathrm{Q_{1}(x,p) = [S]_{x_1}}$ and $\mathrm{Q_{2}(x,p) = [S]_{x_2};}$
        \item $f_{3}(-\infty, p)<\infty.$
\end{itemize}

Here, $S_{i}(x,p) \equiv -4 \pi p \Biggl( \frac{u_{i}}{3}p \frac{\partial f_{i}}{\partial p} + \kappa_{i}\frac{\partial f_{i}}{\partial x} \Biggr)$ is the differential particle flux and $[S]_{x_i}=S_{i}(x,p) \mid_{x>x_i} - S_{i}(x,p) \mid_{x<x_i}$ represents the difference between fluxes on both sides of the shock.

Above we have applied the continuity requirement for particle distribution functions and differential particle fluxes, where $Q_{2}$ represents the preexisting (fossil) mildly relativistic electrons injected for re-acceleration when the merger shock crosses their location on its way to approach the accretion shock. 

Using the above conditions, first we reduce the problem of finding six functions $A_{i}(p),\:B_{i}(p)$ into the coupled system of two first-order differential equations for $A_{1}(p), A_{2}(p), B_{2}(p), B_{3}(p)$  (because $A_{3}(p) = 0 = B_{1}(p)$). Thus, we are finally  able to obtain the ultimate second-order decoupled equation for either $A_{1}(p)$ or $A_{2}(p)$ which has the form of a linear Euler-type equation. The eliminated remaining functions can easily be calculated after solving the latter equation. The general form of a solution for any of these functions has the form

\begin{equation}
B_{i}(p), \: A_{i}(p)= C_{i_1}\cdot p^{\alpha_{1}} + C_{i_2} \cdot p^{\alpha_{2}}
\label{eq:03}
,\end{equation}
where negative spectral indices $\alpha_{1,2}$  are very complicated algebraic functions of many parameters ($x_{1}, \kappa_{2}, u_{2}, r_{1}$ and $r_{2}$). The expression for $\alpha_{1,2}$ is presented below.

\begin{eqnarray}
   \alpha_{1,2} & = & \frac{3}{2}\cdot \Biggl[ \frac{1+l_1\cdot \mathrm{\exp{(k_2\, x_1)}}}{\mathrm{\exp{(k_2\, x_1)}}-1} \nonumber \\
    & & \pm \frac{\sqrt{\mathrm{\exp{(k_2\,x_1)}}\cdot(\mathrm{\exp{(k_2\, x_1)}}+2)\cdot({l_2}^2+l_3)+1}}{\mathrm{\exp{(k_2\, x_1)}}-1} \Biggr]
   \label{eq:04}
,\end{eqnarray}
where
$k_{2}$ is given below after eq. (~\ref{eq:05}) and
\begin{multline*}
    l_1=\frac{r_1 (1-2r_2)+r_2}{({r_1}-1)({r_2}-1)},\\
    l_2=\frac{{r_1}-{r_2}}{({r_1}-1)({r_2}-1)},\\
    l_3=\frac{{r_1}+{r_2}}{({r_1}-1)({r_2}-1)}.\\
\end{multline*}

\begin{figure}[htb]
        \begin{center}
                \includegraphics[totalheight=5.5 cm]{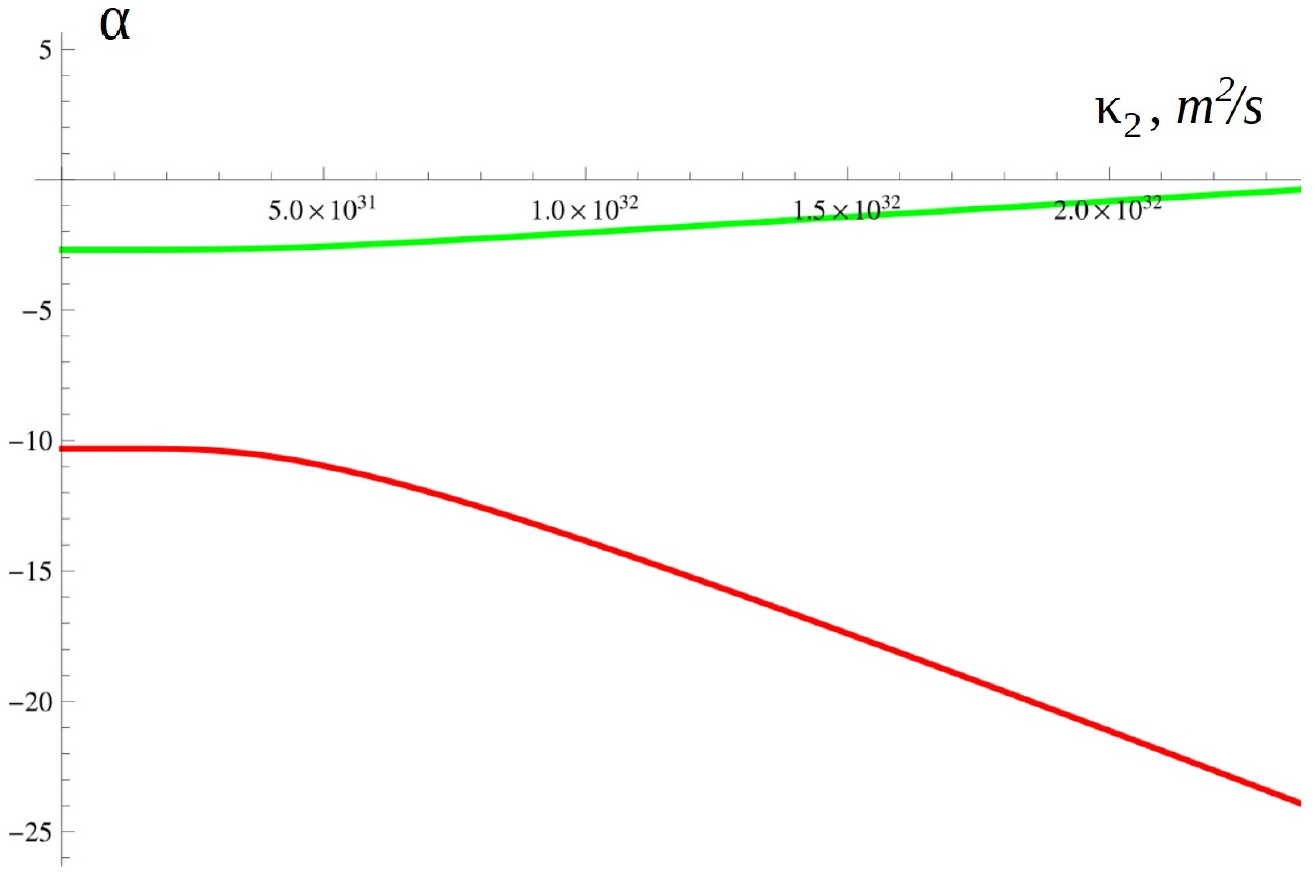}
        \end{center}
        \vspace{-4mm}
        \caption{Dependence of the $\alpha_{i}$ parameters on the diffusion coefficient of the inter-shock region $\kappa_{2}$ (green line: $\alpha_{1}$, red line: $\alpha_{2}$).}
        \label{Figure2}
\end{figure}

A typical functional behavior of these indexes is shown in Fig.~\ref{Figure2}.

Below we restrict our concern to region 2, where particles can potentially participate in acceleration in both shocks. Once the solution of the Euler-type equation for $A_2$ is found and $B_2$ is immediately calculated, one can find the expression for the distribution function $f_2(x, p)$ in the inter-shock region, which is given below:

\begin{eqnarray}
f_{2}(x,p) & = & \frac{1}{k_{1}}\: \Biggl[\mathrm{\exp{[ (k_{2}-k_{1})x_{1}- k_{2} x ]}} \: \Biggl( \xi+ \frac{r_{1}-1}{3 r_{1}} \eta \Biggr)\, \frac{\kappa_{1}}{\kappa_{2}} \: \nonumber \\
& & - \mathrm{\exp{(k_{1}x_{1})}} \Biggl( \xi + \frac{\eta}{3} \Biggr) (r_{1}-1) \Biggr] 
\label{eq:05}
,\end{eqnarray}

where
\begin{multline*}
 k_{i}=u_{i}/\kappa_{i},\\
 \xi =  p^{\alpha_{1}} (C_{1}-G) + p^{\alpha_{2}}(C_{2}+H),\\
 \eta =  \alpha_{1} p^{\alpha_{1}} (C_{1}-G) + \alpha_{2} p^{\alpha_{2}}(C_{2}+H),\\
\end{multline*}
$C_{1}$ and $C_{2}$ are two arbitrary constants,

\begin{eqnarray*}
 G & = & \frac{ \exp^{k_{1}x_{1}} p_{0}^{-(1+\alpha_{1})} (3\, Q_{2}r_{2} + Q_{1} (3 r_{2} \exp^{k_{2}x_{1}}) )}{\kappa_{1} l_{4} \sqrt{ (\exp^{k_{2}x_{1}} + 2) (l_{2}^2 + l_{3})+1}}+\\
  & & + \frac{\alpha_{1} (\exp^{k_{2} x_{1}} - 1) (r_{2}-1)}{\kappa_{1} l_{4} \sqrt{ (\exp^{k_{2}x_{1}} + 2) (l_{2}^2 + l_{3})+1}} ,\\
 H & = & \frac{ \exp^{k_{1}x_{1}} p_{0}^{-(1+\alpha_{2})} (3\, Q_{2}r_{2} + Q_{1} (3 r_{2} \exp^{k_{2}x_{1}} ) )}{\kappa_{1} l_{4} \sqrt{ (\exp^{k_{2}x_{1}} + 2) (l_{2}^2 + l_{3})+1}}+\\
 & & + \frac{\alpha_{1} (\exp^{k_{2} x_{1}} - 1) (r_{2}-1)}{\kappa_{1} l_{4} \sqrt{ (\exp^{k_{2}x_{1}} + 2) (l_{2}^2 + l_{3})+1}} ,\\
 l_{4} & = & \frac{(r_{1}-1)(r_{2}-2)}{r_{1}}.\\
\end{eqnarray*}

The main purpose of this section is to infer information concerning the structure of the solution and to highlight its dependence on important physical parameters.
As seen in Eqs.~(\ref{eq:03})-(\ref{eq:05}), the distribution function is expressed as a combination of two power-law functions whose indexes $\alpha_{i}$ depend on the compression $r_{i}$ of both shocks and the  dimensionless parameter $u_{2} x_{1} / \kappa_{2}$ (which we comment on below).
However, in the case of the combined action of the two shocks, the general shape of $f_2(x, p)$ is also modeled by the additional parameter of relative injection $Q_{2}/Q_{1} $. Therefore, in addition to compression, injection explicitly affects the form of the particle distribution as well, and the spectrum of particles and their synchrotron emission will inherit these influences. Therefore, the interpretation of possible spectral distortions, in particular those found in the lower energy range, should be more distinct and unambiguous in this area because of the lack of significant radiative losses.

In this article, we limit ourselves by analyzing the distribution function of particles as a function of $x$ and $p$. Also, we reveal the impact on its profile made by dimensionless parameters $r_1,\: r_2,\: \kappa_{2}/ \kappa_{1}, \: Q_2 / Q_1$ , and $u_{2} x_{1} / \kappa_{2}$.

The latter parameter affects the distribution function not only through the coefficients in the combination of two power laws but also by the $\alpha_i$ spectral indexes. It is worth noting that the numerical values of $u_{2} x_{1} / \kappa_{2} = L / \lambda_{2}$ , which is equal to the ratio of shock distance over diffusion length. This introduces, on the other hand, a constraint on the permissible $\kappa_2$ value, which ensures a negative value of $\alpha_{1}$  ($\alpha_{2}$ is always negative). The expected negative values of both $\alpha_{i}$ are the result of an asymptotic solution in the case of large $x_{1}$. In order to regain, in the extrapolation process, the appropriate power-law behavior  the shocks must be sufficiently separated, that is, in the case of two independent acceleration processes, one has to obtain power-law spectra $\propto p^{\alpha_{i}}$, where $\alpha_{i} < 0$. It can be shown that for $x_{1} \geq 10$ kpc and for realistic values of $\kappa_{2} \leq 10 ^ {31}\: \mathrm{cm}^{2}/s$, \: $\alpha_{1}$ is negative, which results from the fact that the factor $e^{k_{2} x_{1}} = e^{u_{2} x_{1} / \kappa_{2}} \gg 1$ in eq~(\ref{eq:04}). This in turn is equivalent to $\kappa_{2} / u_{2} \leq x_{1} = L$, and therefore equivalent to the condition of the possible acceleration by both shocks. In other words, depending on the distance between the shocks and the value of $\kappa_{2}$, a certain number of particles can reach the accretion shock, and so both shocks take part in the acceleration processes. We take the value of the inter-shock region width so that an actual double shock system is modeled: if the distance is too great, there is no way the particles can interact with the two shock waves, which simply leads to two separate shock accelerations; if $x_{1}$ is too small,  it is as if only one single shock was operating as accelerator. As a consequence, this can introduce a distortion from a pure power law to a distribution function and subsequently to the electron spectrum.

\section{Results}

In this section we discuss the results obtained on the basis of our model.
Among the three regions appearing in the double-shock model, the inter-shock region is the most interesting: in between two shocks we expect a higher level of turbulence, situated in a collapsing trap of the inter-shock region (see ~\citealt{doi:10.1029/2011JA016669}).

From the theoretical model described in Section 2, we obtain a complicated distribution function $f(x,p)$ depending on many parameters (distance between two shocks, compression ratios, spatial diffusion coefficients, injection values, etc.). Due to the number of parameters that are to be fit, we introduce additional assumptions. As the accretion shock is supposed to be stronger than the merger shock (as shown in \citealt{2015ApJ...812...49H}), here we assume $r_{2}/r_{1}<1$.  In particular, we present the case $r_{1} = 4$ and $r_{2} = 1.5$. Based on the assumptions from \citet{2015ApJ...812...49H}, we take the velocity of the upstream flux to be equal to $\mathrm{u_{2} \approx 10^{6}}$ m/s. We take diffusion coefficients so that $\kappa_{1} > \kappa_{2}$ (due to the increased level of turbulence in region 2) and the source term $Q_{2}/Q_{1} \gg 1$, because the injection of the internal region is expected to be much stronger due to the fact that any nonthermal sources inside the system of merging clusters can provide an additional population of mildly relativistic particles (e.g., AGNs).
It is important to note that the spectrum we expect to see from our analysis is not a power-law `straight' line predicted by DSA, but a bent spectrum, signs of which can be seen both in the observations (see, e.g., Fig. 10 in \citealt{2016A&A...591A.142B}) and in the results of simulations \citep{doi:10.1186/s40623-018-0799-3}. At high frequencies, such an effect may be explained by the influence of the Sunyaev-Zeldovich effect \citep{2016A&A...591A.142B} plus losses and the acceleration processes themselves (e.g., Shock Drift Acceleration mechanism, see \citealt{2011ApJ...742...47M})  on the spectrum. Here, we focus on lower frequency spectral peculiarities. One of the reasons for such peculiarities at low frequencies may be an impact of the accretion shock wave. Below we show the way in which each parameter influences this deviation. We expect the concave spectrum when fitting the parameters $Q_{1},\, Q_{2},\, \kappa_{1},\, \kappa_{2}$ and $x_{1}$.

{Below we quantify or discuss values of $\kappa_{i}, u_i,$ and $ Q_i$ .} Here $\alpha_{i} \equiv \alpha (r_{i}, u_{i}, \kappa_{2}); \: i=1,2$. Spectral indexes are given by eq.4, $\alpha_{2}$ is always negative for any reasonable values of $r_{1}, r_{2}, u_{1}, \kappa_{2}$ whilst $\alpha_{1}$ is negative at particular $x_{1}$ only at a chosen range of $\kappa_{2}$. We take $x_{1} \geq 10 $ kpc, and in our choice of very small $x_1$ we are mostly governed by our interest to consider a system of two shocks shortly before their collision (for a post-collision scenario, see ~\citet{2020MNRAS.498L.130Z, 2020MNRAS.494.4539Z}.

As we consider the inter-shock region to encompass the most interesting physical processes among the other regions, it is the $f_{2}(x, p)$ part of the distribution function that we concentrate on. For simplicity, by default we set $C_{1} = C_{2} = 0$ in cases where no other values of $C_{i}$ are being discussed.
Below, we discuss each parameter in turn, thus dividing our analysis into a number of sections.

\subsection{Source term}

\begin{figure}[htb]
            \begin{center}
                \includegraphics[totalheight=5.5 cm]{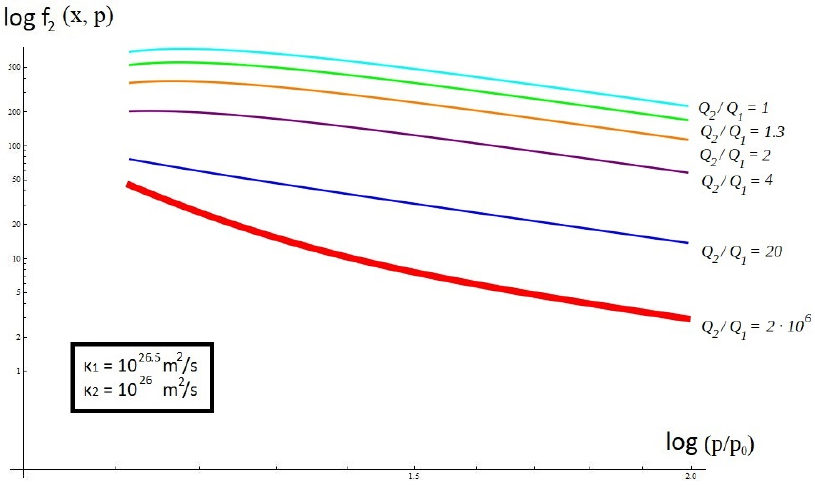}
        \end{center}
        \vspace{-4mm}
        \caption{Dependence of the distribution function of the inter-shock region $f_2(x,p)$ at $x_1\:=\:10\:$kpc, for $Q_2/Q_1\:=\:1,\:1.3,\:2,\:4,\:20,\:2\times10^6\:$on CR momentum.}
        \label{Figure3}
\end{figure}

At closer distances to the galaxy cluster center we assume an additional contribution of particle populations from inside the galaxy cluster (e.g., AGNs). For this reason, the expected relation between the source terms is $Q_{2}/Q_{1} > 1$. Figure~\ref{Figure3} shows that the distribution function $f_{2}(x, p)$ shows a significant dependence on the momentum when $Q_{2}/Q_{1} \gg 1$. As an example, here we compare a particular distribution function $f_2(x,p)$ for several values of the ratio $Q_2/Q_1$ covering the different values of $n_{e,2}$. The series of curves show the whole range of injection parameter $Q_2$ starting from the values characterizing supra-thermal electrons where particle density in a typical ICM environment for different shock locations is equal to $n_{gas,1} \sim 10^{-5} cm ^{-3}$, up to $n_{gas,2} \sim 10^{-2} cm ^{-3}$ (thus injected $n_{e_{th},i}$ fractions are smaller by about $10 ^{4}$ times; see, e.g., ~\citealt{2018NPPP..297...15K}). While for mildly relativistic fossil plasma we expect much more efficient injection resulting in an electron number density of $n_{e,2} \equiv n_{e_{rel},2} \sim 10^{-4} cm^{-3}$.
It is interesting to mention that in cases where $Q_{2}/Q_{1} \leq 4,$ at low energies the shape of the curve may change from concave to convex.

\subsection{Diffusion coefficient}

\begin{figure}[htb]
                \centering
        \begin{subfigure}[b]{\linewidth}
                \includegraphics[totalheight=5.5 cm]{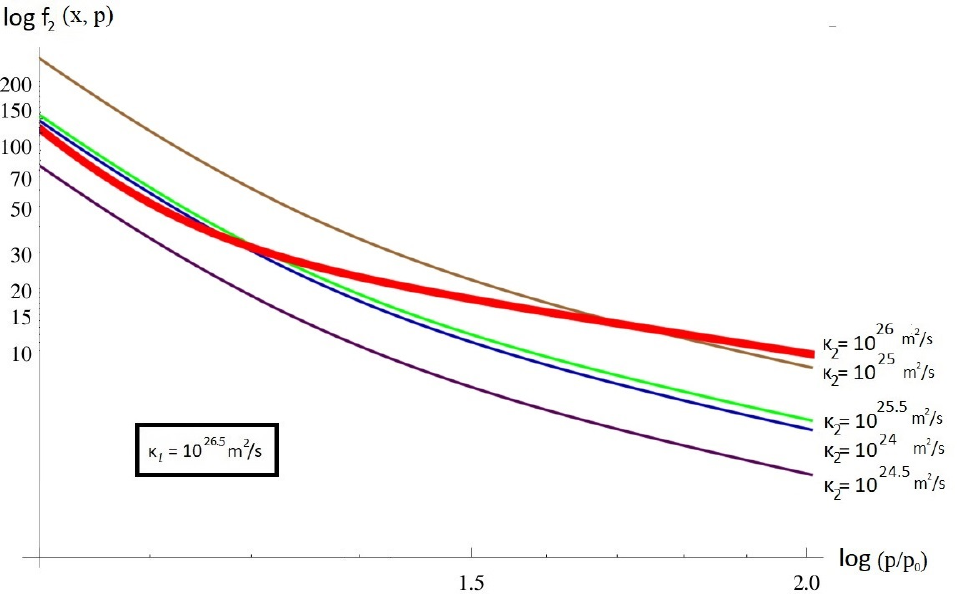}
        \end{subfigure}
        \begin{subfigure}[b]{\linewidth}
                \includegraphics[totalheight=5.5 cm]{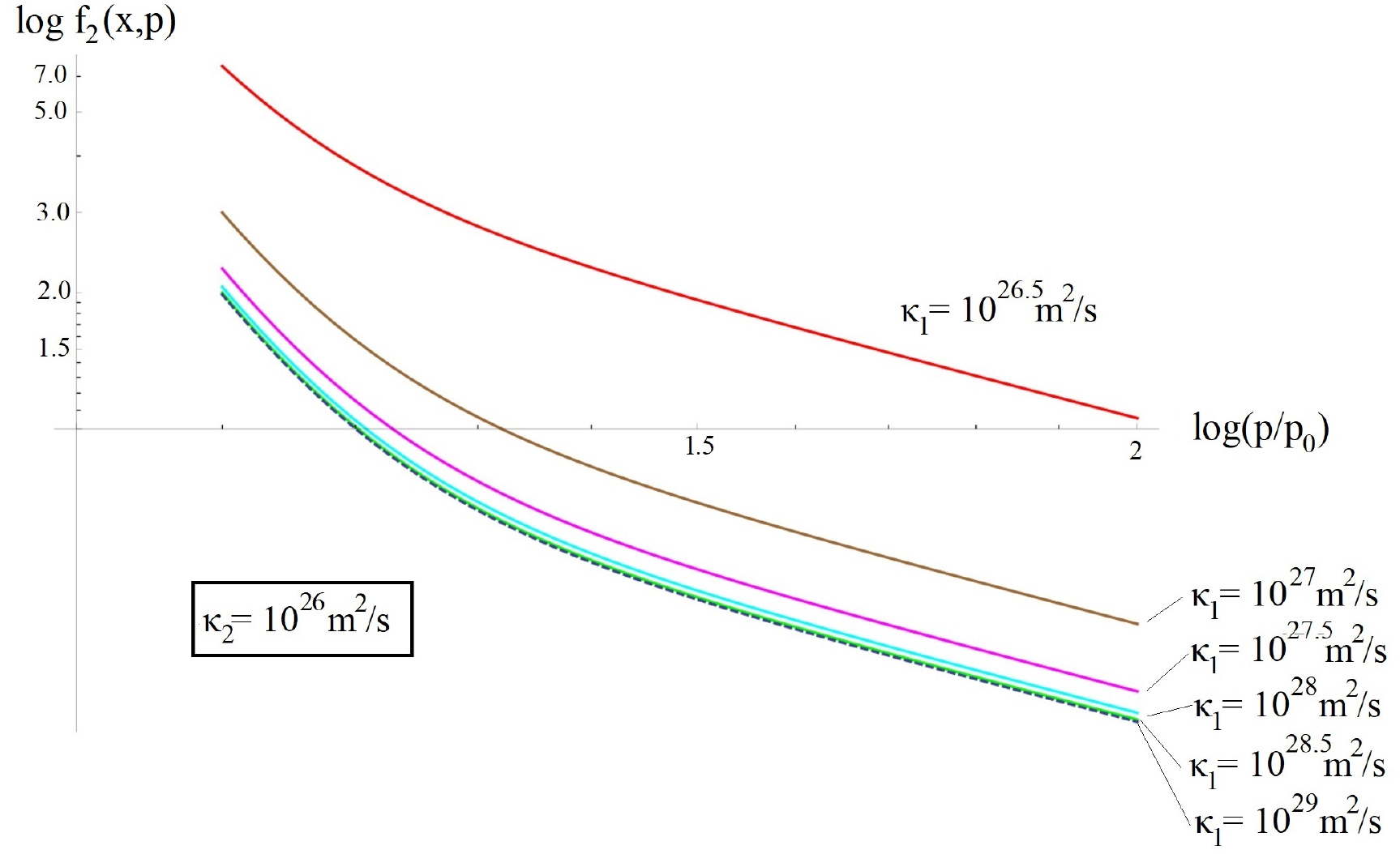}
        \end{subfigure}
        \vspace{-4mm}
        \caption{Dependence of the distribution function on the CR momentum for different diffusion coefficients. \emph{Top panel:} As a function of varying diffusion coefficient of the inter-shock region. \emph{Bottom panel:} As a function of varying diffusion coefficient of the absolute upstream.  We note the dashed ("saturation") curve.}
        \label{Figure4}
        \end{figure}

We set diffusion coefficients as constant throughout the corresponding regions. In this section, two types of plot are shown: $f_{2}(x, p)$ function with respect to $\kappa_{1}$ and $\kappa_{2}$ (see Fig.~\ref{Figure4}).
The values of diffusion coefficients $\kappa_{i}$ are adopted according to the scaling formula (see, e.g., ~\citealt{1997ApJ...487..529B}) which for the particles with $\gamma \propto 10^4$ and for the decreasing intergalactic magnetic field with $x_1$ down to the order of $B \sim 1$nG, one can estimate the value of $\kappa_{1} \sim 3\cdot10^{30} \: cm^2 s^{-1}$ and take the slightly lower value for $\kappa_{2}$.

As we consider an inter-shock region as a collapsing trap, the inequality $\kappa_{1} > \kappa_{2}$ should be fulfilled. Here, the $\kappa_{1} \sim \kappa_{2}$ curve has a different shape from $\kappa_{1} / \kappa_{2} \geq 3.5$ curves (which we represent in bold in Fig.~\ref{Figure4}). The relation $\kappa_{1} / \kappa_{2} \propto 3.2$ shows the most pronounced concave shape. Moreover, the greater the value of $x_{1}$, the larger the distance between the spectral curves with respect to $\kappa_{1}$. Here, $f_{2}(x, p)$ is approximately of the same order as where there is a small distance between the two shocks.

It must also be noted that variation of $\kappa_{1}$ reveals the existence of a limiting (“saturation”) curve of inter-shock distribution function $f_{2}(x, p)$. In the case of very small $x_{1}$, which is of the order of $10$ kpc, the saturation curve occurs at $\kappa_{1} = 10^{28}\: m^{2}/s$ which constrains the $\kappa_{1}$ coefficient range to $10^{26.5}\: m^{2}/s$ to $10^{28}\: m^{2}/s$ at $\kappa_{2} = 10^{26}\: m^{2}/s$ (Fig.~\ref{Figure4}, lower panel).

In order to understand all the effects manifesting in Fig.~\ref{Figure4} it is important to note that we have taken a relatively large value of $\kappa_{2}$ here, which nevertheless still enables acceleration with participation of both shocks. For this chosen value of $\kappa_{2}$, the displayed saturation effect with respect to $\kappa_{1}$ reveals the physical  upper limit for particle acceleration. This means that for larger values of $\kappa_{1}$, the particles meet the upper bound to return to shock 1. Larger values of $\kappa_{1}$, which means larger $\chi$ ($\chi \propto B^2/(\delta B)^2$ in diffusion coefficient), means much weaker scattering and thus less efficient acceleration. In other words, for these larger values of $\kappa_1$ we have to take particle escape into account. The increasing gap (not shown here) in (efficiency) normalization
with shock distance is due to the decreasing  number of particles that take part in acceleration. Here, diffusion length becomes much smaller than the distance that must be reached by particles in order for them to be accelerated in both shocks.  On the contrary, the thicker curve shows that the `double' shock is more efficient at comparable and relatively large values of  both diffusion coefficients. Thus, comparable mean free paths at both sides of shock 1 lead to stronger scattering in the whole region, which leads to a flatter spectrum and greater acceleration efficiency.

\subsection{Position of the probe between the two shocks}

The particular value of $x$ has almost no influence on the profile of the spectrum curve. This means that the position we take in the range of $x_{1} > x > 0$   has very little or no influence on the spectrum.

\subsection{Compression}

\begin{figure}[htb]
                \begin{center}
                \includegraphics[totalheight=4.5 cm]{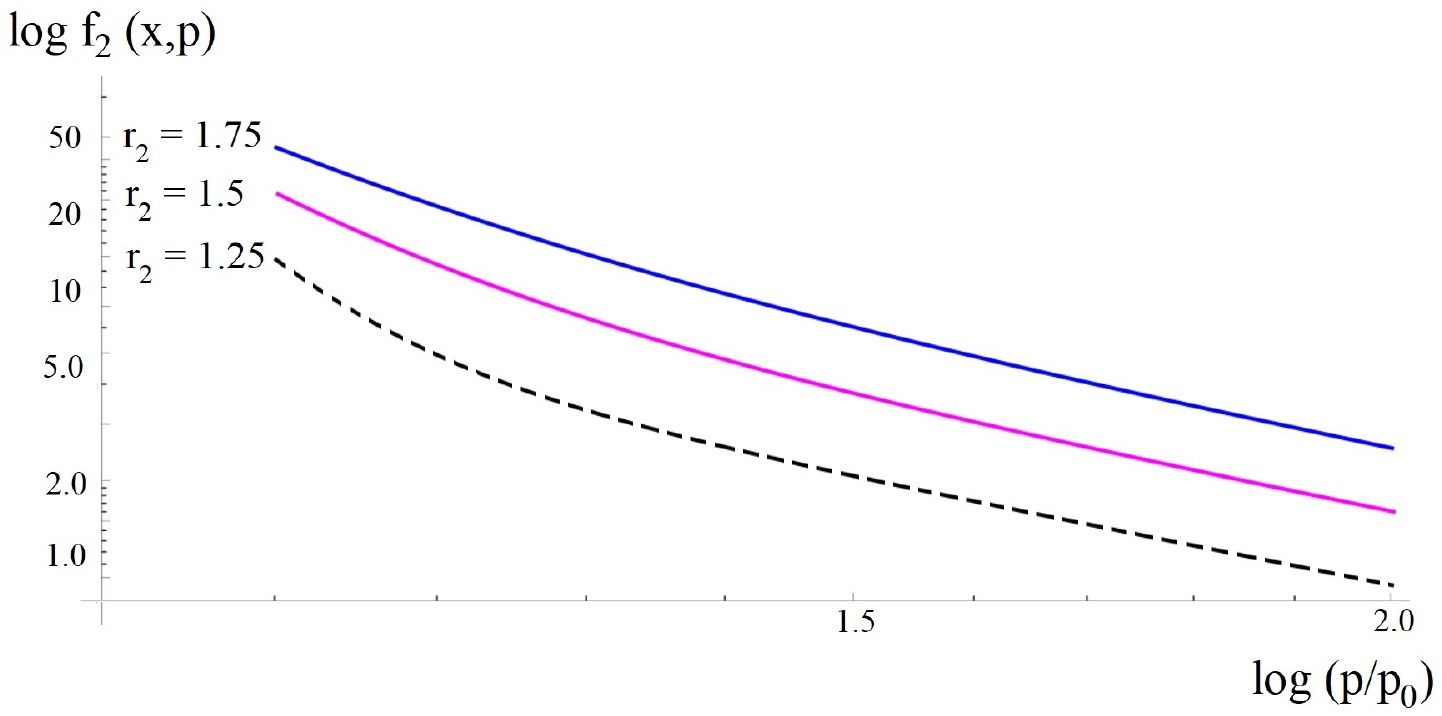}
        \end{center}
        \vspace{-4mm}
        \caption{Dependence of the distribution function on the CR momentum for different values of compression $r_2$ at the position of the second shock, $r_1 = 4$.}
        \label{Figure5}
\end{figure}

\begin{figure}[htb]
            \begin{center}
                \includegraphics[totalheight=5.5 cm]{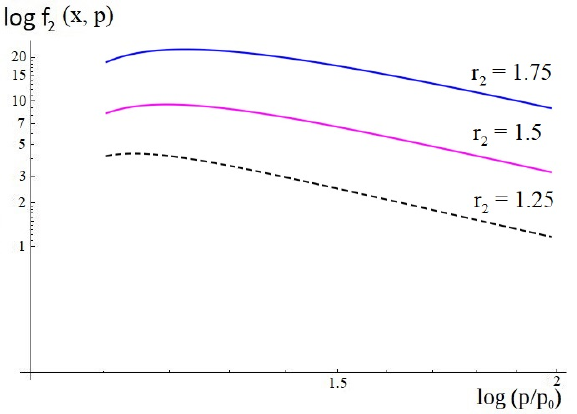}
        \end{center}
        \vspace{-4mm}
        \caption{Dependence of the distribution function on the CR momentum for different values of compression, $r_1$ and $r_2$, at the position of the second shock at $200$ kpc.}
        \label{Figure6}
\end{figure}

\begin{figure}[htb]
            \begin{center}
                \includegraphics[totalheight=5.5 cm]{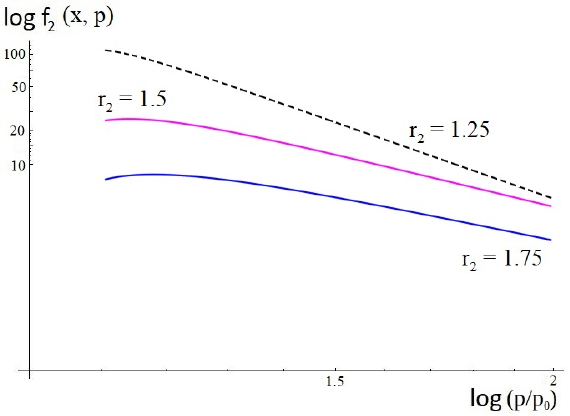}
        \end{center}
        \vspace{-4mm}
        \caption{Dependence of the distribution function on the CR momentum for $r_1\:=\:r_2\:=r$ compression at the position of the second shock at $200$ kpc.}
        \label{Figure7}
\end{figure}

Dependence of $f_{2}(x, p)$ on compression in the inter-shock region $r_{2}$ always reveals an expected bent behavior at $r_{2} < 2$ (Fig.~\ref{Figure5}). Here we set $r_{1} = 4$. It is interesting that in the case of $r_{2} = 1.25,$ the curve has a deeper bending than those with greater values of compression.
Increasing the $x_{1}$ distance up to 200 kpc leads to a visibly different behavior of the distribution function (see Fig.~\ref{Figure6}) with a convex spectrum peculiarity.
As for the case of the equal compressions, $r_{1}=r_{2} = 1.25,1.5,1.75$, the curve corresponding to $r_{1} = r_{2} = 1.25$  is steeper than the others (Fig.~\ref{Figure7}).

Overall, we come to the conclusion that a finite number of variables, such as diffusion coefficients in the absolute upstream and inter-shock regions as well as the injection terms in the same regions, play an ultimate role in the formation of the disturbed spectrum of CRe observed at $200\:-\:300$ MHz.

\newpage

\section{Summary}

Diffuse synchrotron emission radiating from many astrophysical objects sometimes reveals a spectral deviation from a power-law spectrum and we believe that the mechanism behind the representation of this phenomenon should rely on Fermi I acceleration. In the present paper, we concentrate on radio relics, as such objects seem to be a straightforward example of the Fermi I mechanism.

In the current study, we consider a magnetic field  strength in a cluster ICM of around $ 1 \mu$G (see, e.g., Section 4 in \citealt{2019SSRv..215...16V}). In terms of such an assumption, the emission seen at a few hundred MHz is produced by  electrons with energies of a few GeV. We believe that any additional features manifested in the shape of the radio emission spectrum are a reflection of the variation in the momentum slope of the electron spectrum.

The deviations in the flux of radio emission  may happen at a different frequency range. For instance, observations of the Toothbrush and Sausage relics show a clear deviation from a power law observed in the form of a concave shape at low frequencies of around $200-300$ MHz, while at slightly higher frequencies of $1-1.3$ GHz the observations show a bumpy peculiarity (see, e.g., Fig. 10 in \citealt{2016A&A...591A.142B}). The high-frequency spectral steepening in the Sausage and Toothbrush relics may be mainly attributed to losses and could also be contaminated by the Sunyaev-Zeldovich effect \citep{2016A&A...591A.142B}. Conversely, we focus on subtle deviations at about $200-300$ MHz, which to the best of our knowledge have not yet been discussed in literature. We suppose that at the position of a radio relic such a feature is produced by the presence of two shocks. \citet{doi:10.1186/s40623-018-0799-3} present simulations that support our consideration. We believe that the deviation at low frequencies of $\sim200-300$ MHz should mirror the same kind of deviation seen in the electron spectrum. Moreover, we do not expect great losses at the discussed energies: as shown by \citet{2017ApJ...840...42K}, the losses there can be neglected for a field of $\sim1\:\mu$G for low-energy electrons. The concept of two shocks also shows promise for explaining the Mach number and other discrepancies (see, e.g., \citealt{2013MNRAS.433..812O} and \citealt{2016ApJ...818..204V}). Nevertheless, the more far-reaching motive to examine this is to make the DSA model conclusive at low frequencies. We believe that as a consequence, such consideration of the low- and high-frequency parts of the spectrum may  allow the whole radio emission distribution to be properly and consistently described.

In the present paper we study a distribution function of electrons and speculate that the characteristics of its shape could be imprinted in the detected radio spectrum. We analyze the existence of both bumpy or concave features in the model of a double shock wave. Also, we presume that for the two shocks considered in our model (a weaker merger shock and a stronger accretion shock) the origin of injection may differ. We assume that a (internal) merger shock has an additional injection of a CR population from the galaxy cluster volume (AGN influence), while the (outer) accretion shock receives injected particles from the inflow of matter from outside the galaxy cluster. The double shock-wave system suggests the existence of three regions: an absolute upstream region (upstream for accretion shock), an  inter-shock region (which is downstream from the accretion and upstream from the merger shock), and an absolute downstream  region (downstream from the merger shock, instead of upstream and downstream as in the case of one single shock). We expect a higher level of turbulence of the CRs in the inter-shock medium, which means that this region roughly overlaps the relic. The increased turbulence makes this region a kind of collapsing trap \citep{doi:10.1029/2011JA016669} in which the rate of acceleration increases with 
decreasing  distance between the shocks. As the inner merger shock approaches the outer accretion shock, the distance between them decreases making the inter-shock region narrower. According to \citet{doi:10.1029/2011JA016669}, if the second adiabatic invariant remains constant, the parallel momentum of particles trapped between the shocks should increase, increasing the acceleration rate. We believe that, at each region, mechanisms of deviation from the power law are due to different phenomena.

In the present paper we show that the inter-shock region width of about the width of a relic reveals an expected concave spectrum peculiarity, while the increasing distance leads to the appearance of a convex curve. We consider a solution in terms of a transport equation with initial conditions that involve a population of fossil electrons, expressing continuity of both the differential particle flux $S(x, p)$ and the electron distribution function $f(x, p)$ at the positions of the two shocks. In other words, we state that the distribution function or particle flux on both sides of each contact discontinuity are equal to each other, whilst the distribution function far from the galaxy cluster $f(\infty, p)$ is equal to zero and the distribution function close to the cluster center $f(-\infty, p)$ may be of any finite value. In addition to the above-stated consideration, we use a nonstandard representation of the matching conditions representing the flux. This expresses the concept that the additional injection to the system is effectuated in the case where a weak (merger) shock meets fossil populations of electrons on its way to approaching the accretion shock. We believe that even if a weak shock is not visible, it may ignite fossil electrons unevenly distributed in the vicinity of the accretion shock. However, for a bent spectrum to be revealed, both shocks need to be sufficiently close (as we show, down to a distance comparable to the width of a relic). As $Q_{1}$ and $Q_{2}$ are the terms that represent the injection process, the dimensionless relation $\dfrac{Q_{2}}{Q_{1}}$ should be of major importance due to the additional injection of fossil electrons to the supra-thermal part.

We find that the disturbed shape of the spectrum in the inter-shock region mostly depends on the following parameters:

\begin{enumerate}[label=\alph*)]
        \item The ratio of injection. The effect of distortion of the electron spectrum can easily be seen at large ratios of source terms $\dfrac{Q_{2}}{Q_{1}}\propto10^{6}$; this conclusion coincides with our expectations as to the importance of the source terms on the spectrum profile. We consider the difference between $Q_{1}$ and $Q_{2}$ to be the main contributor to spectral deviation at $200-300$ MHz.
        \item Diffusion coefficients of absolute upstream and inter-shock. Although we consider a rather simplistic model in which diffusion coefficients do not depend on energy, even in this case we obtain a distorted spectrum. We see that the influence of $\kappa_{i}$ is very important anyway and conclude that the energy dependence on the diffusion coefficient may be neglected for our current purposes. Different ratios of the absolute upstream  $\kappa_{1}$ and inter-shock  $\kappa_{2}$  diffusion coefficients produce a different behavior of electron spectrum. The spectrum closest to that observed in \citet{2016A&A...591A.142B} and \citet{doi:10.1186/s40623-018-0799-3} is obtained at $\dfrac{\kappa_{1}}{\kappa_{2}}\sim3.2$.
        \item Values of compression and distance between the shocks. Though not as important as injection and diffusion coefficient ratios, and in addition to the distance between the two shocks, the compression also shows an obvious impact on the profile of the CRe spectrum. At inter-shock distances of the order of several tens of kiloparsecs, both at low and strong compression, a convex spectrum peculiarity is observed, while distances of the order of the width of the relic reveal a concave shape. Moreover, spectra corresponding to smaller values of compression are found to be steeper.
\end{enumerate}

As expected, the position of the probe between the two shocks has almost no influence on the profile of the spectrum curve. For this reason, and for the reasons outlined above, we interpret our results as suggesting that the physics behind the injection lead to a much more pronounced effect than the variability in the shock properties.

Our results also appear to suggest that the described discrepancy may be due to the presence of more than one cotemporal discontinuity. As various authors show, the presence of spectral deviations in the Toothbrush and Sausage relics suggest that expansion of DSA theory at low energies is needed. We believe that a double-shock model introduces a fresh approach to the problem. In fact, our model that represents a linear double-shock wave with particles in the inter-shock region that “feel” the presence of both waves, is similar to a single nonlinear shock with a concave power-law spectrum (e.g., Fig. 1 in \citealt{2006MNRAS.371.1251A}). We assume that an  offset of $1$-arcmin (which corresponds to $217$ kpc in the redshift of the Toothbrush cluster) between the position of the merger and the unseen accretion shocks in radio and X-ray bands described in \citet{2013MNRAS.433..812O} is a good illustration of a double-shock model. Another argument in favor of a double-shock model may be the existence of a confusing discrepancy between the Mach number derived from the spectral index and the one estimated from X-ray observations. \citet{2017ICRC...35..283K}  note that a possible explanation to this could lie in the fact that a radio relic may consist of multiple shocks with different strengths. Namely, they state that X-ray observations tend to pick up the parts of shocks with lower Mach numbers and higher kinetic energy flux, while radio emissions come preferentially from the parts of the shocks with higher Mach numbers and higher CR production. Various authors \citep{2017MNRAS.471.1107H,2008ApJ...689.1063S,2009MNRAS.395.1333V} suppose that  the merger shocks responsible for radio relics in galaxy clusters (in LSS formation simulations of the Universe) consist of multiple shocks with different Mach numbers.

\begin{acknowledgements}

We thank to the anonymous reviewer for helpful and insightful comments and to prof. Zdzis{\l}aw Golda for highly valuable assistance in calculating the constituent functions of the distribution function. Mariia Bilinska also appreciates the NASA/IPAC NED operated by Jet Propulsion Laboratory, Caltech, under contract with the NASA, and NASA ADS operated by the Smithsonian Astrophysical Observatory under NASA Cooperative Agreement.
\end{acknowledgements}

\bibliographystyle{aa}
\bibliography{LitBilinska}

\end{document}